\documentclass[a4paper,
               keeplastbox,   % flushend option: not to un-indent last line in References
               ]{jacow}
\usepackage{pdfpages,multirow,ragged2e} %
\makeatletter%
	\ifboolexpr{bool{xetex}}
	 {\renewcommand{\Gin@extensions}{.pdf,%
	                    .png,.jpg,.bmp,.pict,.tif,.psd,.mac,.sga,.tga,.gif,%
	                    .eps,.ps,%
	                    }}{}
\makeatother

\ifboolexpr{bool{xetex} or bool{luatex}} % test for XeTeX/LuaTeX
 {}                                      % input encoding is utf8 by default
 {\usepackage[utf8]{inputenc}}           % switch to utf8

\usepackage[USenglish]{babel}

\ifboolexpr{bool{jacowbiblatex}}%
 {%
  \addbibresource{jacow-test.bib}
  \addbibresource{biblatex-examples.bib}
 }{}
\begin{document}

\title{Upgrade of the SPARC\_LAB LLRF system and recent X-band activities in view of E$\text{u}$PRAXIA@SPARC\_LAB project}

\author{B. Serenellini\thanks{Beatrice.Serenellini@lnf.infn.it}, M. Bellaveglia, F. Cardelli, A. Gallo, G. Latini, L. Piersanti, S. Pioli,
S. Quaglia, \\M. Scampati, G. Scarselletta, S. Tocci, \\INFN - Laboratori Nazionali di Frascati, Via Enrico Fermi 54, 00044 Frascati (RM), Italy}
	
\maketitle

\begin{abstract}
SPARC\_LAB is a high-brightness electron photoinjector dedicated to FEL radiation production and research on novel acceleration techniques. It has been in operation at LNF since 2005. It is composed of a newly designed brazeless 1.6-cell S-band RF gun, two 3 meter long travelling wave S-band accelerating structures, and a 1.4 meter C-band structure that acts as an energy booster. Recently, a plasma interaction chamber was installed to study and optimize beam-driven plasma acceleration schemes. During fall 2023, a major upgrade of the entire low-level RF (LLRF) system will take place to consolidate and improve performance in terms of amplitude, phase resolution, and stability. The original analog S-band and the digital C-band LLRF systems will be replaced by commercial, temperature-stabilized, FPGA-controlled digital LLRF systems manufactured by Instrumentation Technologies. Additionally, the reference generation and distribution will be updated. In parallel with this activity, there is a growing interest in X-band LLRF at LNF due to the EuPRAXIA@SPARC\_LAB project. This project aims to build an FEL user facility driven by an X-band linac at LNF in the coming years. To test X-band RF structures and waveguide components, a high-power X-band test stand named TEX has been installed and recently commissioned. A detailed view of the TEX LLRF system, based on a commercial S-band system with a dedicated up/down-converter stage, will be discussed, along with the limitations of such an approach.
\end{abstract}

\section{SPARC\_LAB test-facility}
\begin{figure}[!htb]
   \centering
   \includegraphics*[width=.9\columnwidth]{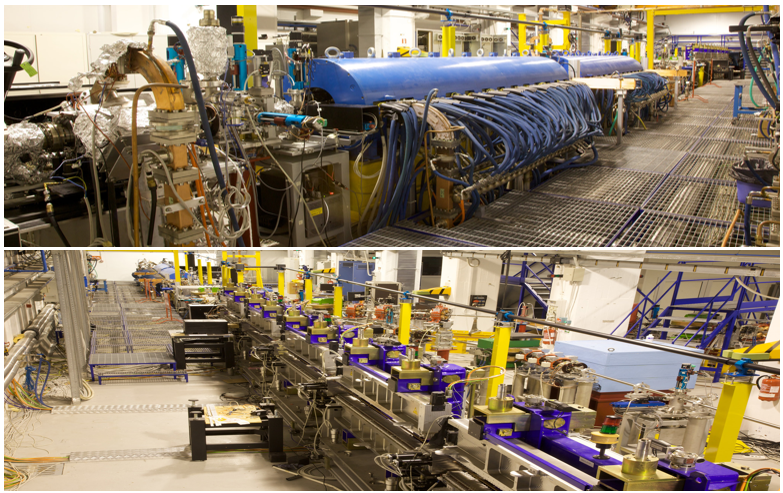}
   \caption{Overview of SPARC\_LAB test facility}
   \label{fig:sparc_lab_photo}
\end{figure}

SPARC\_LAB is a test-facility providing electron bunches with energy up to \SI{170}{MeV} feeding several experimental beamlines, such as SASE and seeded FEL, Thomson back-scattering,THz generation, Plasma focusing and acceleration (Particle WakeField Acceleration-PWFA).\\
The electron source consists of an S-band 1.6 cell BNL/UCLA/SLAC type RF-gun providing \SI{120}{MV/m} peak electric field on the built-in metallic photo-cathode (copper). Electrons are extracted by means of UV laser pulses (wavelength: \SI{266}{nm}, photon energy: \SI{4.66}{eV}) whose shape and duration can be tailored to the needs of different applications and experiments. Particles are accelerated up to \SI{5.3}{MeV} in the gun and then injected into two S-band sections. The first section is also used as RF-compressor by means of the velocity bunching technique. Solenoid coils embedding the first two S-band sections provide additional magnetic focusing and control of the emittance and envelope oscillations, particularly useful in case of velocity bunching operation. A C-band accelerating section is then used as booster, to achieve the nominal kinetic energy. An overview of SPARC\_LAB is shown in Fig.~\ref{fig:sparc_lab_photo} (where the RF photocathode, accelerating sections and the undulator for FEL generation can be identified).
\\
In parallel, in recent years the activities at the LNF are strongly oriented to research very high gradient acceleration with plasma and there has been considerable progress in this field. These novel developments may potentially lead to much more compact and economical accelerator facilities. In this context EuPRAXIA@SPARC\_LAB project was born. In its final configuration it will equip LNF with a multi-disciplinary user-facility, based on a soft X-ray FEL driven by a \SI{1}{GeV} compact X-band RF linac. This fundamental goal will be achieved by using a high gradient plasma accelerator module excited by a driving 0.5 PW-class laser system (FLAME laser). 

\section{Upgrade of the SPARC\_LAB LLRF system}
\subsection{Current LLRF System and Reasons for Updating}
The Reference Master Oscillator (RMO) current in use at SPARC\_LAB, manufacted by Laurin A.G., has coherent outputs at three different frequencies:\\
\begin{Itemize}
    \item  \SI{2856}{MHz}, amplified and distributed to the S-band RF power plants and low level RF (LLRF) front and back-end and is also used to phase-lock the photo-cathode laser and the FLAME laser;\\
    \item  \SI{5712}{MHz}, amplified and distributed to the C-band RF power plants and LLRF front and back-end;\\
    \item \SI{2142}{MHz} ($3/4$ RF S-band), used to down-convert to baseband the signal from a Beam Arrival Monitor (BAM) cavity used for beam timing measurement.\\
\end{Itemize}

Signal distribution is via temperature-compensated coaxial cables and sychronization clients can lock to the distributed reference signal by means of PLLs.\\
At SPARC\_LAB, PFN modulators (from Puls-PlasmaTechnik - PPT) for the S-band and a solid state modulator (from ScandiNova, Sweden) for the C-band are currently used to generate the High Voltage (HV) required for klystrons. The first S-band klystron drives the RF gun and the deflecting cavity used for beam diagnostics. In the waveguide network
there is a circulator in the common path and a variable
attenuator, phase shifter and RF switch in the deflector line.
The second S-band klystron serves the two SLAC-type travelling wave accelerating structures and its RF pulse is compressed with
a SLED type pulse compressor. The C-band plant is dedicated to the constant impedance high gradient (\SI{35}{ MV/m}) structure used as energy booster.\\
The current S-band LLRF system was designed and built at LNF and has been in use since 2006. The front-end is analogue direct conversion based on custom Pulsar Microwave I/Q mixers, has 24 channels and use commercial RF components, is therefore subject to noise and offset limitations. As concerns the digital acquisition of signals, it's performed using commercial cards (made by National Instruments 5105, \SI{60}{MHz}, 12 bit) that have a non-negligible fault rate. \\
The analog back-end employs connectorized RF components (trombone and electronic phase shifters for coarse and fine tuning respectively, variable attenuators and Binary Phase Shift Keying for pulse compressor phase jump). The noise of the front-end is $\sim$ \SI{50}{fs} and no longer meets today's requirements (<\SI{10}{fs}). Furthermore, with a fully analogue system, it is not possible to arbitrarily set the pulse shape and the slow feedbacks against drifts of RF field amplitude and phase have to be performed via control system.\\
The C-band LLRF, instead, is a digital system designed and realized by PSI in the framework of TIARA collaboration. It is made of a 16 channels front-end (>\SI{80}{dB} isolation between channels) that down-converts the RF signals to an IF of \SI{39.667}{MHz} before digitization (16 bit ADC). The back-end has a differential I/Q vector modulator with a bandwidth >\SI{40}{MHz} and an added jitter <\SI{10}{fs}.

\subsection{Updated Layout}
\begin{figure}[!htb]
   \centering
   \includegraphics*[width=.9\columnwidth]{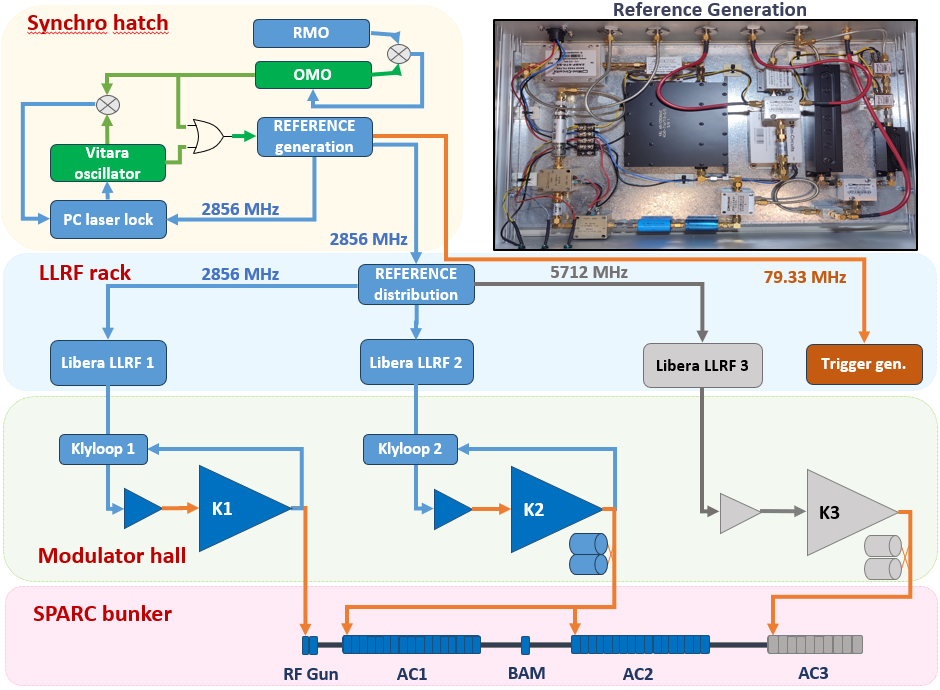}
   \caption{Block diagram of the updated SPARC\_LAB LLRF system and sketch of reference generation}
   \label{fig:Updated SPARC_LAB layout}
\end{figure}

\begin{figure}[!htb]
   \centering
   \includegraphics*[width=.8\columnwidth]{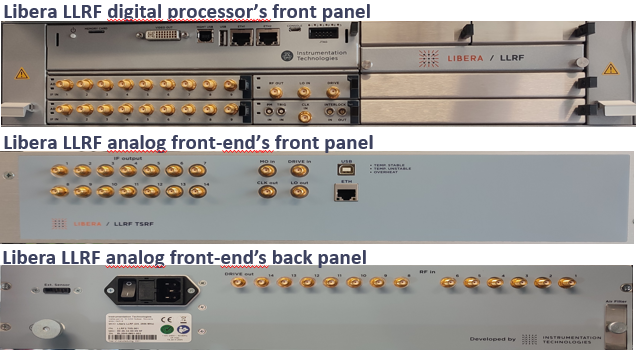}
   \caption{Libera LLRF system: digital processor's front panel (on top), analogue front-end's front panel (in the middle), analogue front-end's back panel (bottom)}
   \label{fig:Libera LLRF}
\end{figure}

The Lazio Regional government recently funded the SABINA (Source of Advanced Beam Imaging for Novel Applications) project for \SI{6.1}{Meuro} for the consolidation of SPARC\_LAB facility. The reference and distribution system will be updated and the LLRF system, both S-band and C-band, will also be completely renewed, standardised and converted to digital. A block diagram of the updated SPARC\_LAB LLRF system and a sketch of custom reference distribution are shown in Fig.~\ref{fig:Updated SPARC_LAB layout}.
\\
Beam timing jitter in a linear accelerator is mainly due to residual phase noise between client oscillators, so the ultimate goal of a synchronisation system is to minimize as much as possible the beam arrival time jitter at certain point of interest. An Optical Master Oscillator (OMO) is tightly locked to the RMO with a relative jitter <\SI{50}{fs} rms. The Vitara laser oscillator (from Coherent) is directly locked to the OMO. The residual in this case is not crucial since all the sub-systems are locked to this same reference, so compared to the current synchronisation scheme (where the laser oscillator is located after the distribution chain), the residual jitter at the RF gun is minimised. To cover all eventualities, a switch was included to choose whether to take the reference from the OMO or the laser oscillator.\\
The hardware for generating the electrical reference and the distribution system were designed and manufactured by the RF LNF group. Commercial components made by Mini Circuits were used (Low Noise Amplifier and power splitter), and custom cavity filters were designed.\\
The new LLRF systems will be furnished by Instrumentation Technologies (Libera LLRF system) and will be able to overcome many of the limitations of current systems. The Libera LLRF system consists of two separate units: an analogue front-end and a digital processor (shown in Fig.~\ref{fig:Libera LLRF}).

Most of the analog signal processing is performed by the components of the Libera LLRF analog front-end unit where LO \& CLOCK and analog front-end boards are located. Both are temperature stabilized in order to avoid the temperature drifts and there are 13 RF inputs available. Down-conversion technique is applied by the analog front-end board to convert the RF signals to Intermediate Frequency (IF = \SI{44.625}{MHz}) signals. The analogue-to-digital conversion of the IF signals is performed by the ADC application board (\SI{14}{bit}, \SI{119}{MHz}, \SI{5}{MHz} BW), and then these digital signals are processed in the FPGA. The analog back-end implements the frequency up-conversion and generates the RF drive signal based on the IF and LO signals. The system has the capability to perform two types of feedback: the first feedback loop operates on the amplitude signal and the second on the phase signal. The digital drive signal is generated in the VM FPGA. Two modes of operation are supported and determine the behavior of the drive signal:\\
\begin{Itemize}
    \item  Open loop: The drive signal is controlled manually with user parameters for open loop amplitude and phase;\\
    \item  Closed loop mode of operation: the drive signal is controlled by independent amplitude and phase feed-back loops where the user sets the target values for amplitude and phase using set-point parameters for amplitude and phase.
\end{Itemize}
In addition, the user can control the shape of the drive’s RF pulse (\SI{16}{bit} DAC, \SI{15}{MHz} BW).
\section{TEX X-band Test-Facility}
\begin{figure}[!htb]
   \centering
   \includegraphics*[width=.7\columnwidth]{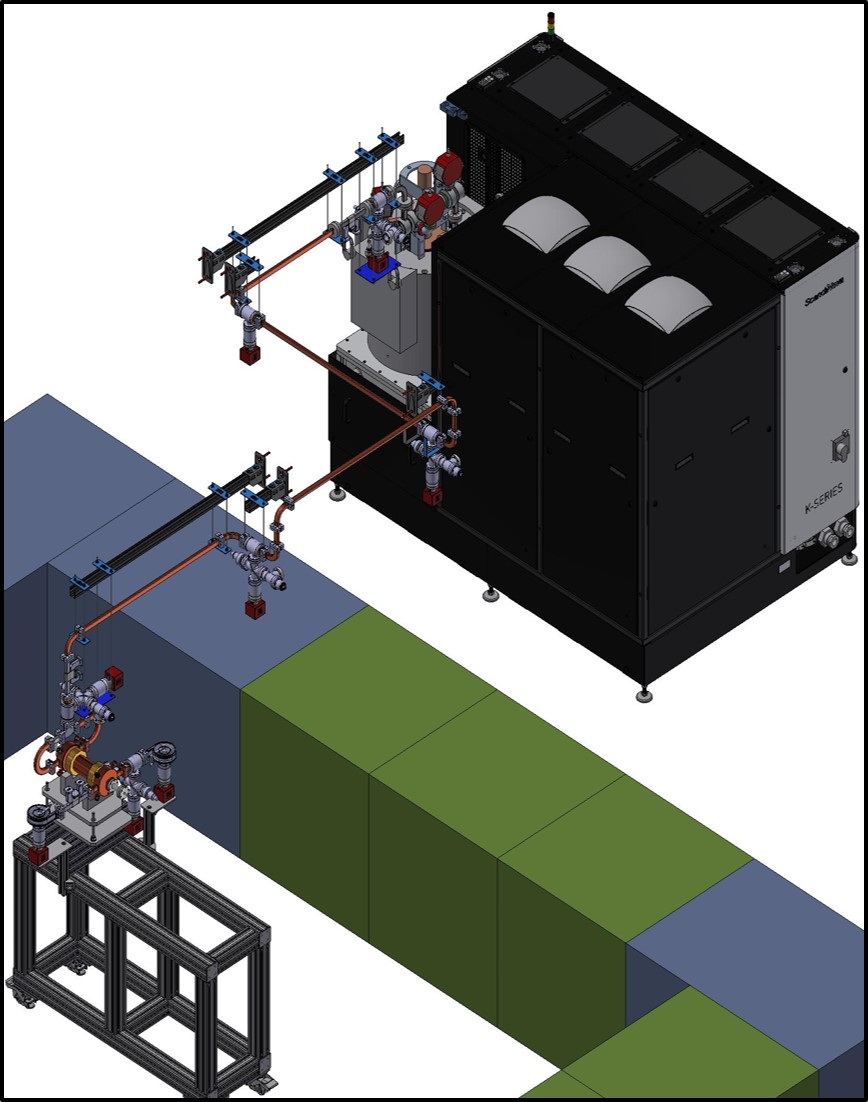}
   \caption{3D CAD drawing of the TEX facility experimental area}
   \label{fig:Drawing of the TEX facility area}
\end{figure}

TEX is a test stand for X-band prototypes of high-gradient accelerating structures and waveguide components created within the LATINO (Laboratory in Advanced Technologies for INnOvation) project funded by the Lazio Regional government. It represents also an R\&D for X-band RF components, LLRF systems, beam diagnostics, vacuum technologies and control system in view of EuPRAXIA@SPARC\_LAB. \\ The RF power plant consists of a K400 450 kV solid state modulator from Scandinova and a VKX8311A 50 MW, 1.5 µs pulsed klystron from CPI LLC (USA). The repetition rate will be at least 50 Hz. A new control room has been realized and equipped, and a concrete shielded bunker has been built to host the devices under test.
In the near future it is planned to install a BOC type RF pulse compressor. A sketch of the TEX area is shown in Fig.~\ref{fig:Drawing of the TEX facility area}, where the modulator cage, the bunker and the waveguide network CAD design is reported.

\section{TEX Facility X-band LLRF System}
\subsection{General Layout}
\begin{figure}[!htb]
   \centering
   \includegraphics*[width=.8\columnwidth]{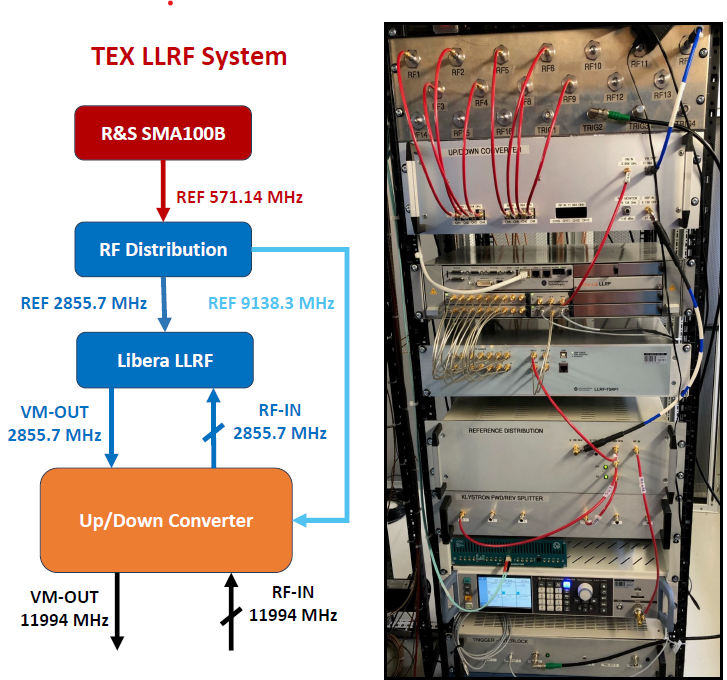}
   \caption{Block diagram of TEX LLRF system (left) and the TEX rack (right)}
  \label{fig:TEX LLRF System}
\end{figure}

Due to the lack of commercial X-band LLRF systems, an S-band Libera system was adapted accordingly. For this purpose, the RF LNF group designed a new hardware that converts the \SI{11.994}{GHz} (European X-band) signal to a frequency of \SI{2.856}{GHz}. Specifically, a reference generation and distribution system able to produce coherent 2.856 GHz S-band and \SI{11.994}{GHz} X-band references, an X-band up/down converter, two custom designed cavity band-pass filters to suppress the signal harmonics and the inter-modulation products in the \SI{9.138}{GHz} reference, were developed. A block diagram of the complete LLRF system is shown in Fig.~\ref{fig:TEX LLRF System}. \\ A picture of the LLRF rack is shown in Fig.~\ref{fig:TEX LLRF System}, where the reference generation and distribution hardware, the Libera front-end and the up/down conversion stage can be identified. As visible from the picture, in addition a splitter stage for klystron forward and reverse power signals, which are needed by the RF mask digitizer (12-bit, \SI{3.2}{GS/s} Wavecatcher) employed for pulse to pulse breakdown detection, a trigger and interlock distribution, to safely inhibit the RF pulses generation whenever the machine protection system detects a possible threat, were developed.

\subsection{Reference Generation and Distribution}
The system RF source is an ultra low phase noise Rohde\&Schwarz synthesizer (SMA100B). It generates a \SI{571.15}{MHz} continuous sine wave, sub-harmonic of both S-band and X-band references, used as baseline for the distribution system, whose output are:\\  
\begin{Itemize}
    \item  Three output at 5 $\times$ RF = \SI{2855.7}{MHz}, one as a reference necessary for the operation of the Libera LLRF and the rest are used for base-band demodulation of klystron forward and reverse signals for machine protection system;\\
    \item  One output at 16 $\times$ RF = \SI{9138.3}{MHz}, necessary for up/down conversion.
\end{Itemize}

\subsection{Up-Down Conversion}
In order to use the Libera LLRF in S-band to process X-band signals arriving from the klystron and accelerating structures, and also to exploit the Libera's vector modulator output, an up/down conversion stage is needed.
Mixers were chosen for the down conversion are the Marki Microwave mixers MT3-0113HCQG. These mixers can be operated with a LO power as high as \SI{24}{dBm} and have an high linearity with respect to RF power. This effect is particularly beneficial, since the Libera front-end dynamic range extends up to \SI{+10}{dBm}. A Marki Microwave M2B0218HP RF mixer is used in the up conversion stage. This choice has been driven by the high LO power (up to \SI{22}{dBm}) and high linearity that allows to fully exploit the Libera \SI{12}{dBm} vector modulator output at \SI{2.856}{GHz}. For the purposes of suppressing inter-modulation products on the IF port of mixers, the RF LNF group also designed and built customised band-pass filters. 

\section{System Performance}
\begin{figure}[!htb]
   \centering
   \includegraphics*[width=.9\columnwidth]{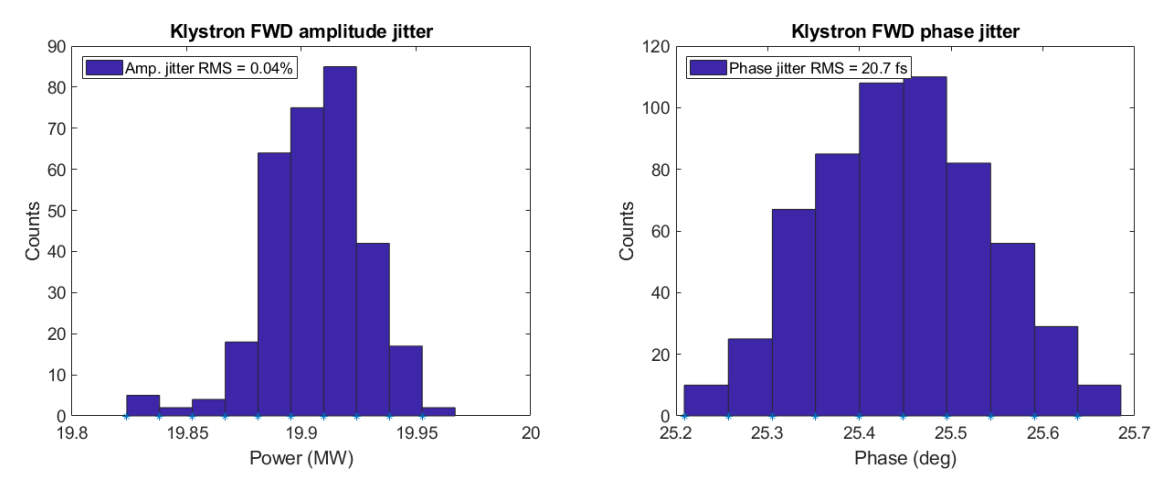}
   \caption{Measured amplitude and phase jitter of klystron output forward power. The measurements refer to \SI{20}{MW}, \SI{300}{ns} at \SI{50}{Hz} repetition rate RF pulses}
   \label{fig:jitter}
\end{figure}

During the SAT, some preliminary measurements have been carried out aiming at characterizing the amplitude and phase jitter on the klystron output power, which is one of the most important parameters to qualify the stability of the power source. In Fig.~\ref{fig:jitter} the outcome for \SI{20}{MW}, \SI{300}{ns} pulse amplitude (left) and phase (right) stability measurements are shown, while Tab. ~\ref{tab:meas jitter} summarizes the results. Very promising jitter values have been obtained: 0.04 \% and \SI{20.7}{fs} rms for amplitude and phase respectively. These performance are compatible with state of the art solid state power sources, and represent a strong groundwork for the future X-band linac facility EuPRAXIA@SPARC\_LAB.

\begin{table}[!hbt]
   \centering
   \caption{RF jitters preliminary measurement}
   \begin{tabular}{lcc}
       \toprule
       \textbf{Signal} & \textbf{Amp. Jitter (\%)}                      & \textbf{Phase Added Jitter (deg)} \\
       \midrule
        kly frw         & 0.04    & 0.0894 (\SI{20.7}{fs})        \\ %[3pt]
       \bottomrule
   \end{tabular}
   \label{tab:meas jitter}
\end{table}
\section{CONCLUSION}
The SPARC\_LAB facility was presented with a focus on the synchronisation system and the LLRF systems. Subsequently, the reasons why this systems are obsolete in terms of phase jitter and therefore the need for an upgrade were explained. The updated LLRF (Libera LLRF) system will be digital and will overcome most of the limitations of the current system. The Libera system has already been used in the TEX (test stand for X-band prototypes) facility, where an up/down conversion system was used to exploit an S-band Libera. Finally, preliminary measurements of amplitude and phase jitter on the klystron output power were presented.

\ifboolexpr{bool{jacowbiblatex}}%
	{\printbibliography}%
	{%
	% "biblatex" is not used, go the "manual" way
	
	%\begin{thebibliography}{99}   % Use for  10-99  references
	
}

\end{document}